\documentstyle[prb,aps,twocolumn,epsf]{revtex} 
\begin{document}

\title{Magnetic behaviour of quasi-one-dimensional oxides, 
Ca$_3$Co$_{1+x}$Mn$_{1-x}$O$_6$}

\author{S. Rayaprol, Kausik Sengupta  and  E.V. Sampathkumaran\footnote{Corresponding author: sampath$@$tifr.res.in}}
 
\address{Tata Institute of Fundamental Research, Homi Bhabha Road, 
Mumbai - 400 005, INDIA.}

\maketitle

\begin{abstract} 
{The results of ac and dc magnetization  and heat capacity 
measurements on  the oxides, Ca$_3$Co$_{1+x}$Mn$_{1-x}$O$_6$, forming in  a K$_4$CdCl$_6$-derived rhombohedral quasi-one-dimensional crystal structure, are reported. As far as  Ca$_3$Co$_2$O$_6$ is concerned, the results reveal truly complex nature of the two magnetic transitions, identified to set in  at 24 and 12 K in the previous literature. However, partial replacement of Co by Mn apparently results in a long magnetic ordering of an antiferromagnetic type (below 13 and 18 K for x= 0.0 and 0.25 respectively), instead of spin-glass freezing in spite of the fact that there is Co-Mn disorder; in addition, interestingly there are hysteretic spin reorientation effects as revealed by isothermal magnetization behavior.}
\end{abstract}
\vskip0.5cm
$^*$E-mail address: sampath@tifr.res.in
\vskip1cm

%\newpage
\maketitle

The compounds of the type, (Sr, Ca)$_3$ABO$_6$ 
(A, B= a metallic ion, magnetic or nonmagnetic), crystallizing in the 
K$_4$CdCl$_6$ (rhomhohedral) derived structure\cite{1} (space group R${\bar 3}$c) have started attracting the attention of   
condensed matter physicists in recent years (see, for instances, Refs. 2-13), as the 
structure is of  a pseudo-one-dimensional-type, characterized by the presence of magnetic chains of A and B ions running 
along c-direction arranged hexagonally  forming a triangular lattice. 
These chains are  separated by Sr (or Ca) ions. Among these, the compound, Ca$_3$Co$_2$O$_6$, is interesting in many ways. Among this class of oxides, this is the only compound in which both A (octahedral oxygen coordination) and B (trigonal prismatic oxygen coordination) sites are occupied by the same metallic ion. The interchain distance is 5.24 $\AA$, whereas the intrachain Co-Co distance is about 2.6 $\AA$ and thus this is a good candidate for  quasi-one-dimensional magnetism.   It has been proposed that this compound serves as a rare example for "partially disordered antiferromagnetic structure (PDA)" in the sense that, for an intermediate temperature (T) range (12-24 K), 2/3 of the ferromagnetic-Co chains are antiferromagnetically coupled to each other, whereas the rest remain incoherent; as the T is lowered below 12 K,  according to  original reports,\cite{3} these incoherent chains undergo spin-freezing with the application of a magnetic field (H) inducing ferrimagnetic structure.   However, the appearance of prominent plateaus\cite{3,5} in the plot of isothermal magnetization (M) in the range of 12 to 24 K is in favor of ferrimagnetic ordering for this intermediate T range as well.  
 Subsequent neutron diffraction studies,\cite{4} revealed the  existence of long range ferrimagnetic correlation not only above 12 K, but also below 12 K.    Though there are such confusing reports about the precise T/H region over which ferrimagnetic structure appears, it is certain that  this compound is believed to be the first example for ferrimagnetism in a planar Ising triangular lattice, noting that each chain acts effectively as a single spin.\cite{3}  
In addition, there is a superparamagnetic-like large frequency ($\nu$) dependence\cite{5} of real part ($\chi$$\prime$) of ac susceptibility ($\chi$), which does not categorise this into "conventional spin-glasses".  Though it is known that Co at the A-site (Co-I) is bigger in size compared to that at the B-site (Co-II), there has been an ambiguity about the valency of Co,  as (i) the question whether orbital angular momentum is quenched remains unanswered, and (ii)  it is difficult to consistently interpret the magnetic moment values obtained from M and neutron diffraction data.\cite{3,7}  A recent article on band structure calculations\cite{13} reports that the oxidation state of Co at both the sites is 3 with the one at the A-site in a low-spin state and the other at the B-site in a high-spin state.   Many other interesting observations\cite{8,9} have also been made on this compound.  However, it is obvious from the above discussion that the understanding of this compound is thus still at a premature stage and more experimental data need to be collected which will eventually contribute to better understanding of this compound. 

Here, we report the results of ac $\chi$ measurements taken in the presence of different H as well as of heat-capacity (C) measurements on this compound, while dc M measurements have also been measured to compare with the literature data. The results reveal new interesting features.  During the course of our investigations of this class of compounds,\cite{11,12} we also came across another article recently,\cite{14}  which reports the synthesis  of Mn substituted  oxides, Ca$_3$Co$_{1+x}$Mn$_{1-x}$O$_6$ (x= 0.0 and 0.25), the magnetic properties of which are ambiguous. Keeping this in mind, we considered it worthwhile to probe how the properties of this Co compound get modified by partial Mn substitution. We therefore subjected the oxides with the compositions, x= 0.0, 0.25 and 1.0, to  careful M and heat-capacity (C) measurements, the results of which are reported in this article.

Polycrystalline specimens of Ca$_3$Co$_{1+x}$Mn$_{1-x}$O$_6$ (x= 0.0, 0.25 and 1.0) were prepared by a conventional solid state method as described in Ref. 14. Proper proportions of high purity ($>$ 99.9 $\%$) CaCO$_3$, 
Co$_3$O$_4$ and MnO$_2$  were thoroughly mixed in an agate mortar and calcined in 
air at 1273 K for one day and then at 1275 K for x= 1.0 and at 1475 K for x= 0.0 and 0.25 respectively for another day with an intermediate grinding. The speciens were subsequently subjected to a further heat-treatment in  oxygen atmosphere at 1275 K for 4h with subsequent slow cooling. The samples were then characterized by x-ray diffraction (Cu K$_{\alpha}$) and found to form in single phase.  The x-ray diffraction patterns (see Fig. 1) are found to be in excellent 
agreement with those reported in Ref. 14. Dc $\chi$ 
measurements (1.8 - 300 K) in the presence of various H  and 
isothermal M measurements at several temperatures were performed employing a 
superconducting quantum interference device (Quantum Design) as well as a vibrating sample magnetometer (VSM) (Oxford Instruments). The same Quantum Design  
magnetometer was employed to measure ac $\chi$ (2-200 K) in a  
$\nu$ range of 1 - 1000 Hz in an ac field of 1 Oe in zero dc H for all samples and in the presence of 
two different dc magnetic fields (H= 10 and 50 kOe) for x= 1.0. The C 
measurements (2 - 60 K) were performed by  semi-adiabatic heat-pulse method 
employing a home-built calorimeter.

We first discuss the results on Ca$_3$Co$_2$O$_6$. As shown in figure 2a, the plot of inverse $\chi$ versus T measured in a H of 5 kOe is nearly linear in the range 130 to 300 K. The effective moment ($\mu_{eff}$) and the paramagnetic Curie temperature ($\theta_p$) obtained from the slope of this plot is found to be 5.1 $\mu_B$ per formula unit and 30 K respectively. There is a sudden increase in $\chi$ below 24 K, as though there is a ferromagnetic transition. In the data for the zero-field-cooled (ZFC) state of the specimen, as the T is lowered, this feature is  followed by a peak at 12 K indicating the existence of another (antiferro-magnetic-like) transition at this temperature. The ZFC-FC $\chi$ curves (Fig. 2b) for H= 5 kOe bifurcate below 12 K only, but not at 24 K, which establishes that the 24K-transition can not be of a spin-glass-type. These observations are in broad agreement with those known in the literature. With respect to dc $\chi$(T) behavior, we make the following new observations:  The bifurcation of ZFC-FC curves, obtained in a  lower field of 100 Oe does not begin at 12 K (see Fig. 2b), but apparently occurs at a lower temperature only, say near 7 K; also there is no peaking of $\chi$ at 12 K in the ZFC plot for this field; it may be noted that there is an additional shoulder at the same T (7 K) in ZFC data of 5 kOe. The appearance of the 7K-feature thus depends on T and H cycling history of the specimen. A careful look at the curves for 1 kOe in Fig. 2 of Ref. 3 reveals the existence of an upturn of $\chi$ at nearly the same temperature, though this observation was not emphasized by those authors.  These features may point to the existence of another  transition at 7 K.      

In Fig. 3, we show the isothermal M behavior at selected temperatures for the above compound. There is a plateau in the plots (with increasing H) at about 1/3 of the saturation value (approximately equal to the value at the highest field measured) at about 20 kOe for intermediate T ranges, say at  5, 8 and 20 K. The plots at these temperatures are less hysteretic compared to that at 1.8 K. The saturation moment values (say, at 120 kOe) are nearly the same for all T below 24 K. These observations are  in agreement in with those reported in the literature.\cite{3,5} The new observations we would like to emphasize from  our data are: (i) The plots are non-hysteretic at 20 K as well as at 30K, in contrast to those noted in Ref. 3 for 20 K; (ii) For T= 5 K, there are additional structures (as though there are more plateaus) while reversing H to zero, as indicated by vertical arrows in this figure; (iii) At 1.8 K, there are many steps with increasing H, which is however reduced with the reversal of the direction of  H. It should be mentioned that such finer steps could be observed only with VSM, as this instrument enables to collect the data at very close intervals of H. These  results clearly suggest that there are  many energetically close magnetic structures, the stability of which are extremely sensitive to the H and/or T cycling. 

We now turn to the ac $\chi$ behaviour (Fig. 4), both real ($\chi$$\prime$) and imaginary ($\chi$$\prime$$\prime$)  parts. There is a peak at 12 K for the lowest $\nu$ in $\chi$$\prime$(T) plot. There is a similar feature in  $\chi$$\prime$$\prime$ near the same T (10 to 12 K). These peaks undergo a shift to a higher T with increasing $\nu$. These are all characteristics of spin glasses. What is surprising is that, as noted earlier,\cite{5} the magnitude of the shift of the peak temperature is too high (about 5 K) to classify this compound as a canonical spin-glass. This "superparamagnetic-like" behavior was believed\cite{5} to arise from many energetically close magnetic structures inferred from isothermal M behavior. But, we are hesitant to support this arguement, considering that another isostructural compound, Ca$_3$CoRhO$_6$, with a more  dramatic frequency dependent ac $\chi$ behavior\cite{11} does not show  such  multiple steps in isothermal M.   It is interesting to see that the ac $\chi$ attains almost zero value for the highest frequency (1000 Hz) below 10 K, as though the M is no longer able to follow the excitation field, thereby yielding zero average value.\cite{5} We have obtained additional information on the basis of field-dependent ac $\chi$ measurements, the results of which are also shown in Fig. 4. For the purpose of direct comparison of  the H-dependence of the intensities of the peaks, we have plotted all the curves in the same scale. It is obvious that an application of 10 kOe completely suppresses the peak at 12 K and a new $\nu$  independent weak peak appears around 24 K  in $\chi$$\prime$, however without any worthwhile feature in the imaginary part. This implies that this feature arises from long range magnetic ordering at 24 K, which apparently is masked in the zero field data by the tail of the 12K-peak at the higher temperature side.  Interestingly, further increase of H, say to 50 kOe, restores the 12K-peak with noticeable frequency dependence (though with less intensity) not only in $\chi$$\prime$(T), but also in $\chi$$\prime$$\prime$(T); $\chi$$\prime$ however falls gradually with increasing T without any feature at 24 K.  This "oscillatory" field-dependent ac $\chi$ intensity behaviour is a new observation and quite fascinating. Careful neutron diffraction measurements in the presence of H will be quite rewarding.

In Fig. 5, we report the results of C measurements (in zero H). There is a distinct peak in C(T) at 24 K with a large jump of about 3.8 J/mol K. This finding is consistent with the fact that long range magnetic ordering sets in at this temperature. However, C is found to decrease monotonically down to 2 K with further decrease of T, thereby establishing that the magnetic transitions at 12 and 7 K are not  of a long range type. 

We now present the results on the Mn substituted samples. In the case of Ca$_3$CoMnO$_6$, the plot of inverse $\chi$ versus T is linear down to about 50 K, below which there is apparently a small deviation from linearity (Fig. 6a). The values of $\mu_{eff}$ and $\theta_p$ obtained from this linear region are about 6.0$\mu_B$ and  -45 K respectively.  The negative sign of $\theta_p$  indicates dominant antiferromagnetic correlations.     Further lowering of T show a broad shoulder at about 13 K, suggesting the existence of a magnetic transition at this temperature in conformity with Ref. 14. However, in our data, this feature is followed by an another upturn below 7 K (see Fig. 6) and it is not clear whether this originates from traces of impurities appearing in the x-ray diffraction pattern. C data (Fig. 5) also show a corresponding anomaly around 13 K and the rounded nature of the peak in C(T) may indicate  that  the magnetic correlations may not proceed through the entire crystal, which may be responsible for the absence of magnetic reflections in the neutron data.\cite{14} We do not find any difference between ZFC $\chi$ and FC $\chi$; the peak temperature in $\chi$$\prime$ (at the magnetic transition) does not exhibit any $\nu$ dependence and the $\chi$$\prime$$\prime$(T) is featureless (and hence not plotted here). These observations, somewhat different from what are observed for Ca$_3$Co$_2$O$_6$, establish that the  magnetism of this compound  is relatively simpler,  that is, of a long range type. This conclusion resolves that the absence of  clear-cut magnetic reflections in the neutron diffraction data is not due to spin-glass-type of magnetic ordering.\cite{14} Isothermal M behaviour (Fig. 6d) is quite revealing. The M(H) plot at 30 K is practically linear (at least up to 60 kOe) with a weak curvature at higher fields. This tendency persists at 10 K as well, without any evidence for hysteresis, from which one can conclude that the magnetic ordering can not be of a ferromagnetic or spin-glass-type. Interestingly, as T is lowered to 5 K, there is a dramatic change in the M behavior: M is a non-linear function of H even in the relatively low  field range, say, till 20 kOe and at a field of about 30 kOe, there is a meta-magnetic-like transition followed by a linear variation above 80 kOe; in addition, hysteretic behavior is observed for the intermediate field range. Similar features are observed at 1.6 K with a more pronounced hysteretic behavior. All these results suggest that the zero-field magnetic structure is of an antiferromagnetic-type and hysteretic M(H) behavior at low temperatures  signals that  the metamagnetic transition could be of a first order, broadened    by possible crystallographic (Mn/Co) disorder. It is interesting that the inferred crystallographic disorder however does not result in spin-glass-like anomalies. 
 
	We  show the results of magnetic investigations for Ca$_3$Co$_{1.25}$Mn$_{0.75}$O$_6$ in Fig. 7. It may be remarked that we have not attempted to synthesize more compositions, rich in Mn, as Ref. 14 claims that such compositions do not form. From the comparison of the results shown in Figs. 6 and 7, it is obvious that the magnetic behavior of this compound is qualitatively similar to that of Ca$_3$CoMnO$_6$, except that the long range (antiferro)magnetic ordering sets in at about 18 K, without any evidence for additional low temperature transition. This is confirmed by the appearance of a corresponding feature in the C data (Fig. 5). There is a weak upturn below 7 K if dc $\chi$ is measured with H= 100 Oe (Fig. 7b), and this is suppressed in H= 5 kOe data implying thereby that this upturn could be due to traces of magnetic impurities, at least for this case.  The low temperature field-induced spin-reorientation transition (Fig. 7d) is relatively broadened and occurs at a higher field range, around 60 to 80 kOe, however maintaining hysteretic behavior, while compared with Ca$_3$CoMnO$_6$.  The dc $\chi$ obeys Curie-Weiss law with a small deviation below about 100 K and the values of $\mu_{eff}$ and $\theta_p$ obtained from the high-T linear region (5.7$\mu$$_B$ and -24 K respectively) are in good agreement with those reported in Ref. 14. It may also be remarked that we do not see a bifurcation of ZFC-FC $\chi$ curves in the T range of  measurement, similar to the behavior of Ca$_3$CoMnO$_6$ - in contrast to the observation in Ref. 14 - consistent with our conclusion that the magnetic ordering is not of a glassy-type.        

To conclude, we have reported detailed ac and dc $\chi$ and C behavior of spin-chain compounds, Ca$_3$Co$_{1+x}$Mn$_{1-x}$O$_6$. We report many new findings on Ca$_3$Co$_2$O$_6$.  The results reveal that there are apparently many magnetic structures, which are energetically close, the stability of which depends on thermal and magnetic field cycling history of the specimens. Interesting features are also seen in the field-dependent ac $\chi$(T) plots.    Partial replacement of Co by Mn however drives this system towards long range antiferromagnetic ordering, with  field-induced spin-reorientation effects, interestingly,  with an hysteretic effect for intermediate field range. It is fascinating that chemically induced disorder by the substitution of Mn for Co does not favor spin-glass-like freezing, whereas the stoichiometric compound, Ca$_3$Co$_2$O$_6$ exhibits "exotic" spin-glass-like features, thereby revealing that this compound is truly a novel magnetic material. It will be rewarding to subject this compound to more  investigations by other methods.        

The authors would like to thank Kartik K Iyer for his help while 
performing the experiments.

%\newpage

%\newpage

\begin{figure}
\centerline{\epsfxsize=7cm{\epsffile{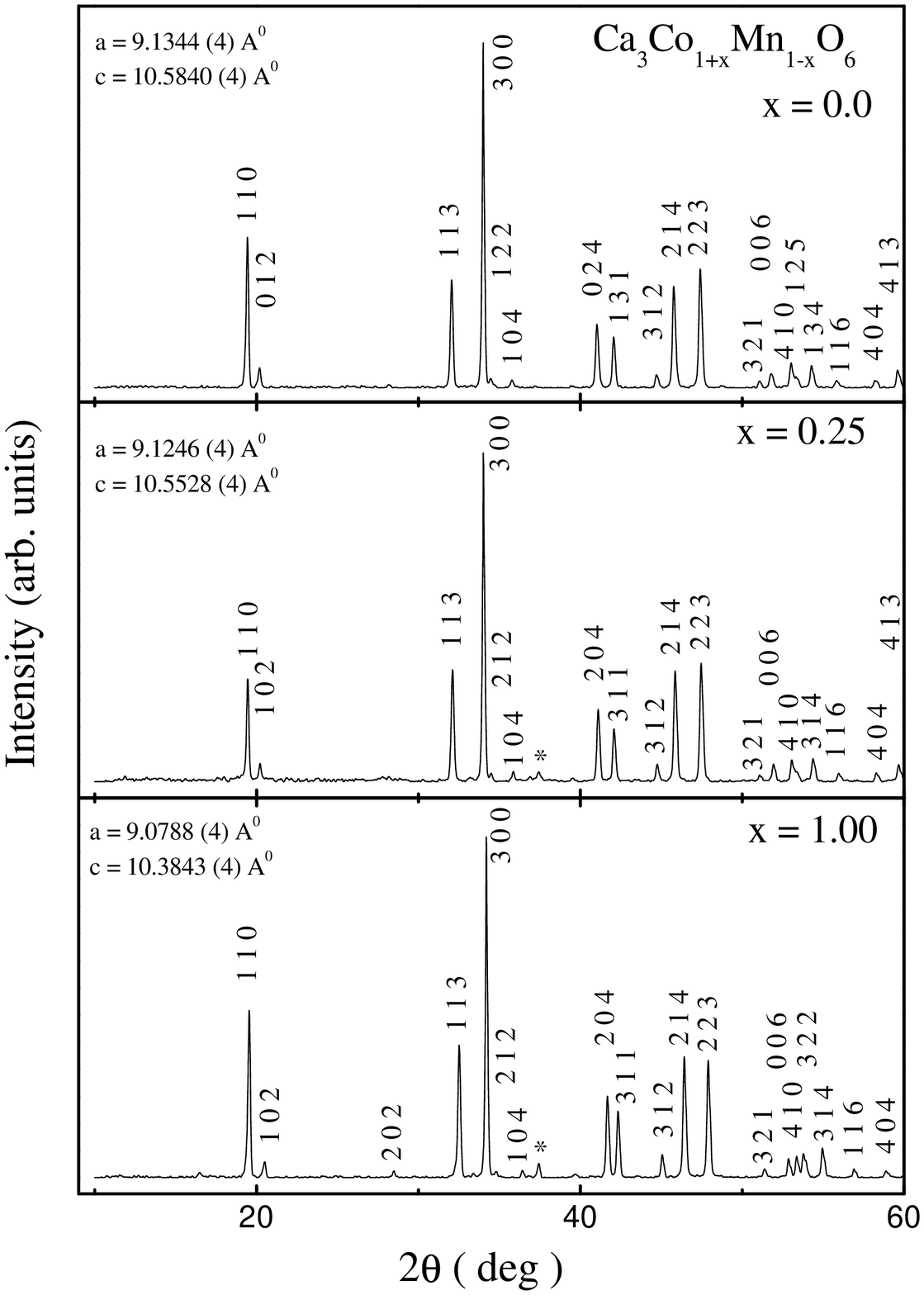}}}
\caption{X-ray diffraction patterns (Cu K$_\alpha$) of the oxides, Ca$_3$Co$_{1+x}$Mn$_{1-x}$O$_6$. The lattice parameters are included in the figures. Asterisks mark  unidentifiable weak peaks.}
\end{figure}

\begin{figure}
\centerline{\epsfxsize=5cm{\epsffile{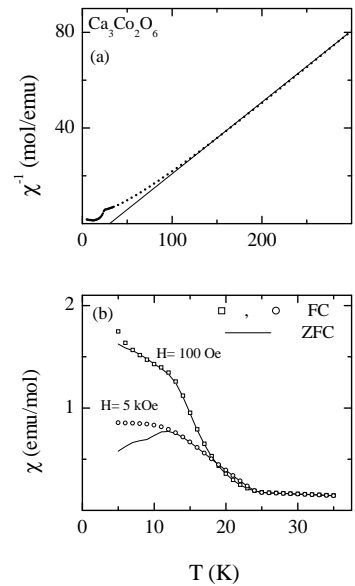}}}
\caption{(a) Inverse dc magnetic susceptibility ($\chi$) as a function of temperature 
(T) for zero-field-cooled Ca$_3$Co$_2$O$_6$ measured in the presence of an external magnetic 
field (H) 5 kOe. The line represent high-T linear region. (b) Dc $\chi$ taken in the presence of 5 kOe and 100 Oe for zero-field-cooled (ZFC) and field-cooled (FC) states of the specimens are compared. }
\end{figure}

\begin{figure}
\centerline{\epsfxsize=6cm{\epsffile{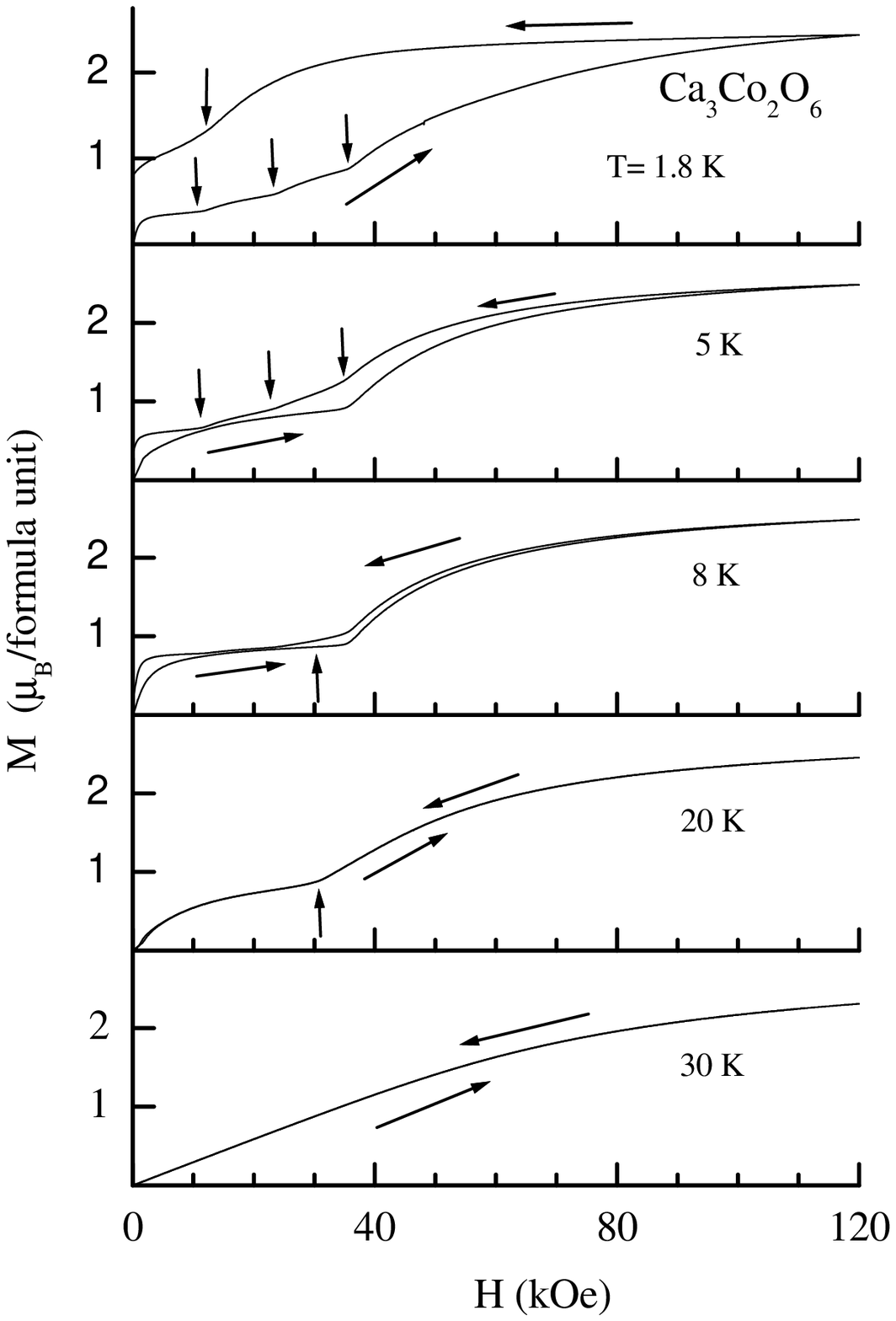}}}
\caption{Isothermal magnetization (M) as a function of magnetic field (H) for Ca$_3$Co$_2$O$_6$ at various temperatures. Vertical arrows mark the steps in the plots, while other arrows mark the direction of variation of the field.}
\end{figure} 
%\vskip3cm

\begin{figure}
\centerline{\epsfxsize=8cm{\epsffile{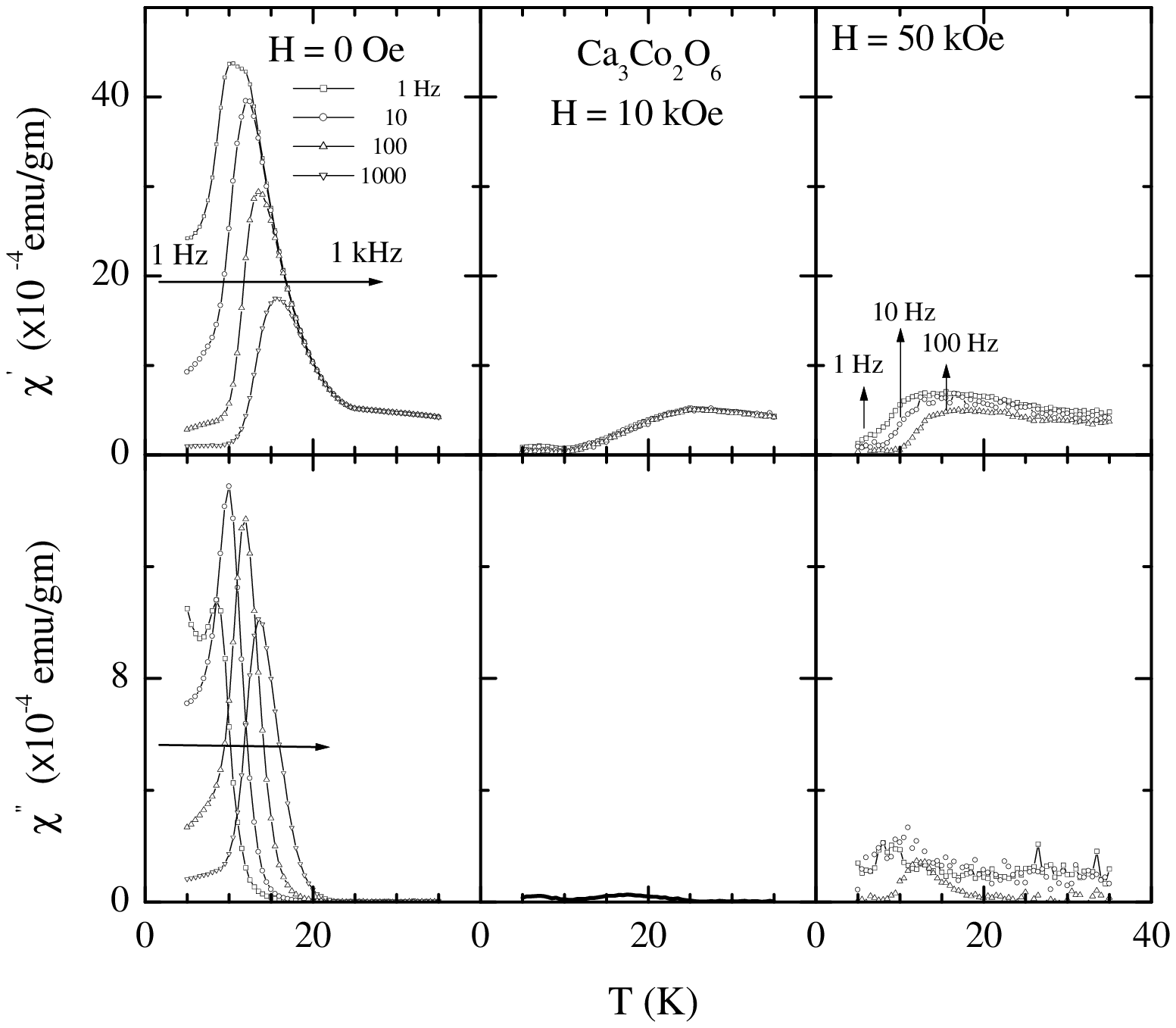}}}
\caption{Ac susceptibility as a function of temperature for various 
frequencies in the absence and in the presence of external dc magnetic 
fields (10 and 50 kOe) for Ca$_3$Co$_2$O$_6$. $\chi\prime$ and $\chi\prime\prime$ refer to real and imaginary 
parts. For H= 10 kOe, the values at various $\nu$ overlap. In the zero-field data, the peaks shift in the direction of arrows with increasing frequency.}
\end{figure}

\begin{figure}
\centerline{\epsfxsize=5cm{\epsffile{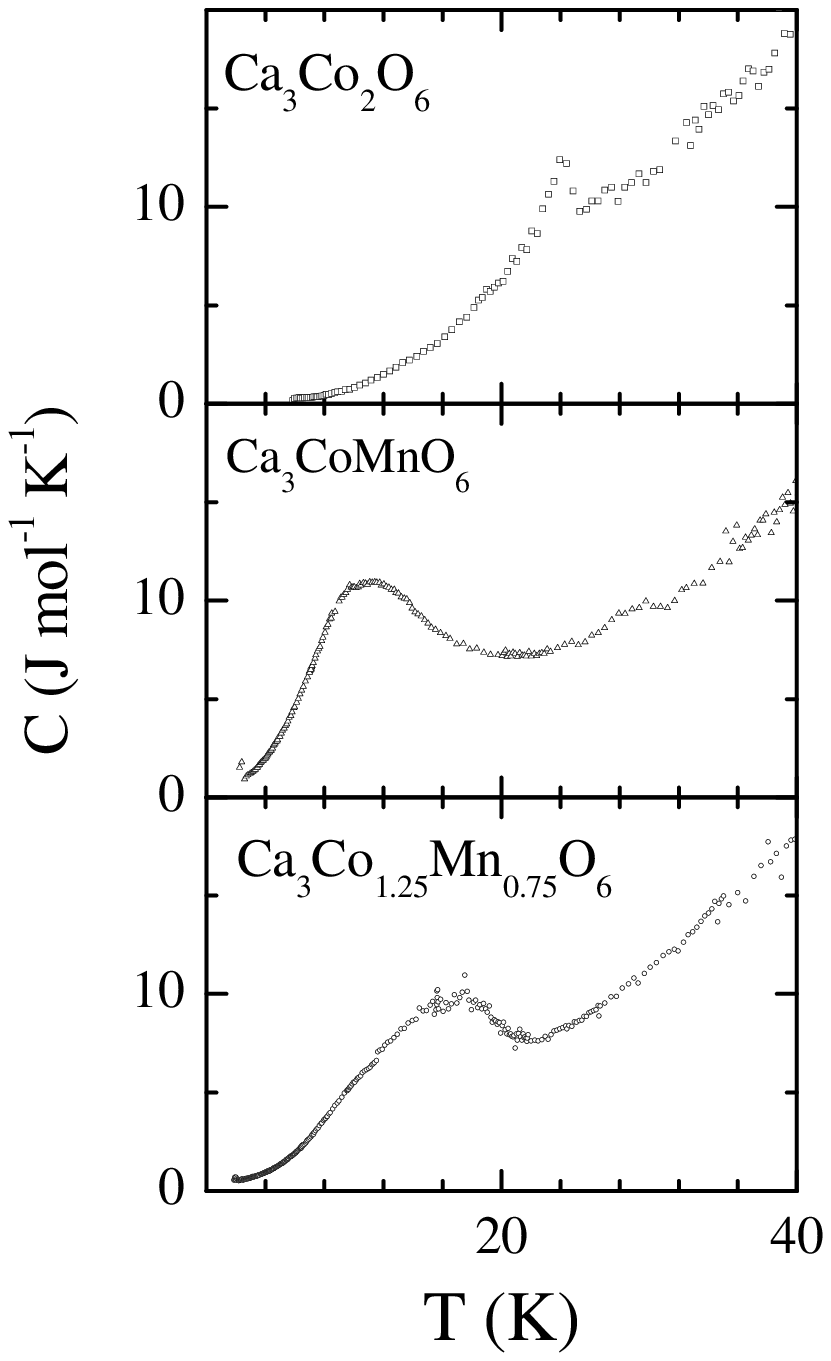}}}
\caption{Heat capacity (C) as a function of temperature (T) for 
Ca$_3$Co$_{1+x}$Mn$_{1-x}$O$_6$ (x= 0.0, 0.25 and 1.0).}
\end{figure}

%\vskip1cm
\begin{figure}
\centerline{\epsfxsize=8cm{\epsffile{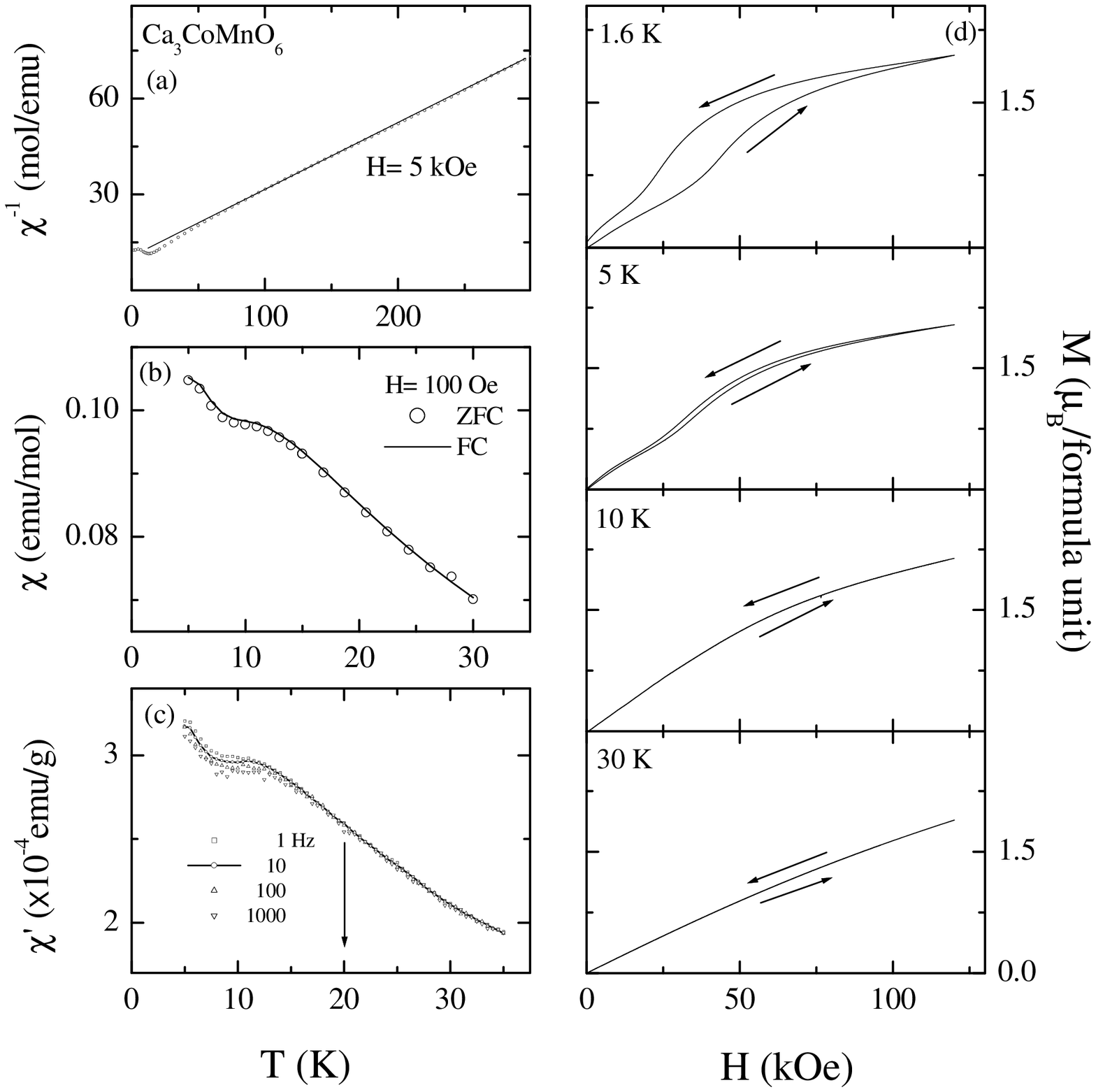}}}
\caption{ For Ca$_3$CoMnO$_6$, (a) inverse dc susceptibility ($\chi$) as a function of temperature (T); a continuous line represents high-T linear region; (b) Dc $\chi$ as a function of T measured in a field of 100 Oe for the ZFC and FC state of the specimen; (c) ac $\chi$ (real part) as a function of T at various frequencies; and (d) isothermal magnetization (M) as a function of magnetic field (H) at selected temperatures with the arrows marking the direction in which the H is varied.}
\end{figure}

\begin{figure}
\centerline{\epsfxsize=8cm{\epsffile{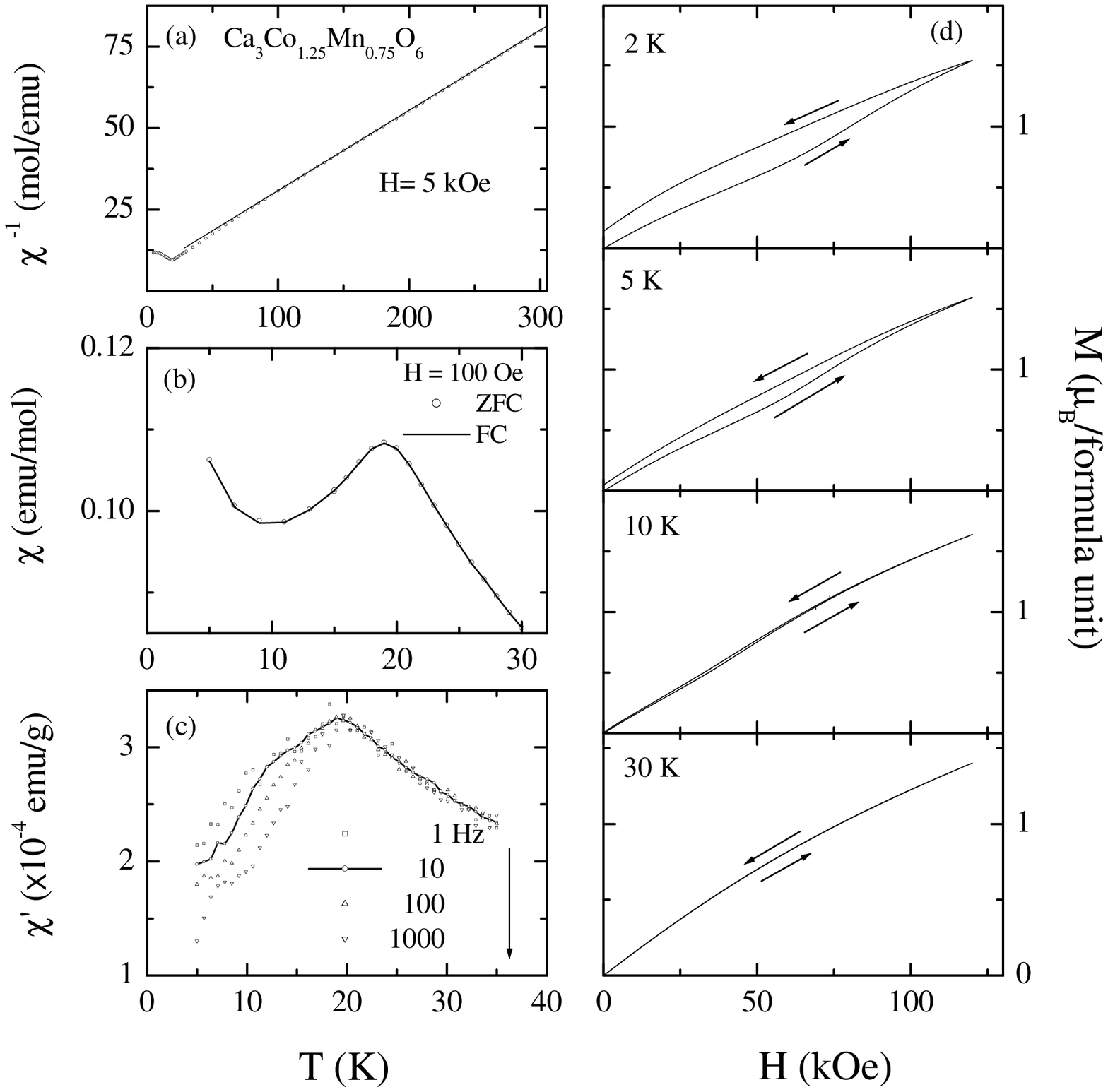}}}
\caption{For Ca$_3$Co$_{1.25}$Mn$_{0.75}$O$_6$, (a) inverse dc susceptibility ($\chi$) as a function of temperature (T); a continuous line represents high-T linear region; (b) Dc $\chi$ as a function of T measured in a field of 100 Oe  for the ZFC and FC state of the specimen; (c) ac $\chi$ (real  part) as a function of T at various frequencies; and (d) isothermal magnetization (M) as a function of magnetic field (H) at selected temperatures with the arrows marking the direction in which the H is varied.}
\end{figure}


\begin{references}

\bibitem{1} J. Darriet, F. Grasset and P.D. Battle, Mat. Res. Bull. {\bf32}, 139 (1997); C.A. Moore, E.J. Cussen and P.D. Battle, J. Solid State Chem., {\bf153}, 254 (2000); T.N. Nguyen and H.C. zur Loye, J. Solid State Chem. {\bf117}, 
300 (1995).
\bibitem{2}H. Kageyama, K. Yoshimura and K. Kosuge, J. Solid State Chem. 
{\bf140}, 14      (1998).
\bibitem{3}H. Kageyama, K. Yoshimura, K. Kosuge, H. Mitamura and T. Goto, J. 
Phys. Soc. Jpn. {\bf66}, 1607 (1997).
\bibitem{4}H. Kageyama, K. Yoshimura, K. Kosuge, X. Xu and S. Kawano, J. Phys. Soc. Jpn. {\bf67}, 357 (1998). 

\bibitem{5}A. Maignon, C. Michel, A.C. Masset, C. Martin and B. Raveau, Eur. 
Phys. J. B {\bf15}, 657 (2000). 

\bibitem{6}H. Fjellvag, E. Gulbrandsen, S. Aasland, A. Olsen aaand B.C. Hauback, J. Solid State Chem., {\bf124}, 190 (1996). 

\bibitem{7}S. Aasland, H. Fjellvag and B. Hauback, Solid State Chem., {\bf101}, 187 (1997). 

\bibitem{8}H. Kageyama, S. Kawasaki, K. Mibu, M. Takano, K. Yoshimura and K. 
Kosuge, Phys. Rev. Lett. {\bf79}, 3258 (1997).


\bibitem{9}B. Raquet, M.N. Baibich, J.M. Broto, H. Rakoto, S. Lambert and A. 
Maignan, Phys. Rev. B {\bf65}, 104442 (2002).

\bibitem{10}S. Niitaka, K. Yoshimura, K. Kosuge, M. Nishi and K. Kakurai, 
Phys. Rev. Lett. {\bf87}, 177202 (2001).

\bibitem{11}E.V. Sampathkumaran and Asad Niazi, Phys. Rev. B {\bf65}, 180401 
(2002). 
\bibitem{12}M. Mahesh Kumar and E.V. Sampathkumaran, Solid State 
Commun. {\bf114}, 643 (2000); Asad Niazi, P.L. Paulose and E.V. Sampathkumaran, Phys. Rev. 
Lett. {\bf88}, 107202 (2002); P.L. Paulose, M. Mahesh Kumar, Subham 
Majumdar and E.V. Sampathkumaran, J. Phys. Condens. Matter {\bf12}, 8889 
(2000); Asad Niazi, E.V. Sampathkumaran, P.L. Paulose, D. Eckert, A. 
Handstein, and K.-H. M\"uller, Solid State Commun. {\bf120}, 11 (2001); 
Phys. Rev. B {\bf65}, 064418 (2002); Asad Niazi, P.L. Paulose, E.V. 
Sampathkumaran, Ute Ch. Rodewald, and Jeitschko, Pramana - J. Phys. {\bf58}, 
1069 (2002); S. Rayaprol, Kausik Sengupta and E.V. Sampathkumaran, Phys. Rev. B {\bf67}, 180404(R) (2003); Kausik Sengupta, S. Rayaprol, Kartik K. Iyer and E.V. Sampathkumaran, Phys. Rev. B, in press.   

\bibitem{13}M.-H. Whangbo, D. Dai, H.J. Koo and S. Tobic, Solid State Commun. {\bf125}, 413 (2003). 

\bibitem{14}V.G. Zubkov, G.V. Bazuev, A.P. Tyutyunnik and I.F. Berger, J. Solid State Chem., {\bf160}, 293 (2001). 




\end{references}
\end{document}